\documentclass[12pt]{iopart}
\usepackage{iopams}
\begin{document}
\title{Numerical Relativity Beyond Astrophysics}

\author{David Garfinkle}

\address{Dept. of Physics, Oakland University, Rochester, MI 48309, USA}
\address{Michigan Center for Theoretical Physics, Randall Laboratory of Physics, University of Michigan, Ann Arbor, MI 48109-1120, USA}

\ead{garfinkl@oakland.edu}

\begin{abstract}

Though the main applications of computer simulations in relativity are to astrophysical systems such as black holes and neutron stars, nonetheless there are important applications of numerical methods to the investigation of general relativity as a fundamental theory of the nature of space and time.  This paper gives an overview of some of these applications.  In particular we cover (i) investigations of the properties of spacetime singularities such as those that occur in the interior of black holes and in big bang cosmology.  (ii) investigations of critical behavior at the threshold of black hole formation in gravitational collapse.  (iii) investigations inspired by string theory, in particular analogs of black holes in more than 4 spacetime dimensions and gravitational collapse in spacetimes with a negative cosmological constant.

\end{abstract}

\maketitle

\section{Introduction}

	General relativity is Einstein's theory of gravity.  For weak gravitational fields the predictions of general relativity are well approximated by those of Newtonian gravity, so the main applications of general relativity are to those astrophysical situations where gravity is strong: neutron stars, black holes, and the big bang.  However, general relativity is also a fundamental theory of the properties of space and time.  Therefore, there are interesting questions in general relativity that have nothing to do with astrophysics, including (1) the nature of the spacetime singularities formed in gravitational collapse and occurring at the big bang; (2) the process of black hole evaporation and the quantum state of the radiation that is left after evaporation is complete; (3) the prospects for combining general relativity with quantum mechanics to form a quantum theory of gravity.  
	
Numerical relativity means computer simulations applied to general relativity.  One uses computer simulations when other methods are not sufficient, in particular when no exact solution or perturbative expansion can be expected to give accurate results.  One of the main astrophysical applications of numerical relativity has been to simulate the collision of two black holes including the gravitational waves produced by such collisions.\cite{frans,manuella,baker,saul}  (Related work involves simulating the collision of two neutron stars\cite{shibata,stu,luis}, or of a black hole and a neutron star\cite{shibata2,shapiro2}).  Numerical relativity has also been used to study the properties of neutron stars and of the core collapse supernovae\cite{ott,tony} that can lead to the formation of either neutron stars or black holes.
	
On the non-astrophysical side, numerical relativity is used to study the properties of spacetime singularities.  Theorems due to Hawking and Penrose and others\cite{hawking} tell us that singularities form in very general circumstances once gravity gets sufficiently strong.  But these theorems tell us almost nothing about the properties of these singularities.  There are various conjectures about the properties of such singularities, and in particular about the rate at which the gravitational field blows up.  Computer simulations are performed of gravitational collapse leading to spacetime singularities to evaluate which of the conjectures are correct.
	
One interesting phenomenon discovered using numerical relativity is critical gravitational collapse\cite{choptuik}: the properties of the threshold of black hole formation.  If one chooses initial data with a free parameter  that measures how strongly gravitating the matter is at the initial time, then weakly gravitating initial data will lead to dispersion of the matter, while strongly gravitating data lead to the formation of a black hole.  There will therefore be some critical value of the parameter that just barely leads to black hole formation.  Matt Choptuik performed detailed simulations of the collapse of a self-gravitating scalar field and found something interesting: there is a scaling relation between the mass of the black hole formed and the amount that the parameter differs from its critical value.  This is reminiscent of the scaling found near phase transitions in condensed matter physics, but now with the black hole mass playing the role of the order parameter.  Furthermore, Choptuik found that the critical solution (the one that results from the critical value of the parameter) is discretely self similar: in a certain amount of time the scalar field evolves to the same profile with the scale of space reduced.  An infinite number of these cycles takes place in a finite amount of time, with the result that the critical solution forms a spacetime singularity that is not hidden inside a black hole.  Since Choptuik's simulations critical collapse has been studied for many different types of matter\cite{gundlachlr}.  For simplicity, Choptuik performed his simulations in spherical symmetry, but since then the phenomenon has also been studied in axisymmetry.  One challenging project for the future is to simulate critical collapse in the absence of any symmetry at all.  
	
One field that has led to many new applications of general relativity is string theory.  In particular, the low energy limit of string theory is essentially general relativity with a somewhat exotic collection of matter fields, and in more than 4 spacetime dimensions.  This has led to renewed interest in the Einstein field equations in the context of low energy string theory.  Much of this work has involved finding closed form exact solutions\cite{garyandtoby}, though some numerical work has also been done to find these solutions\cite{wiseman}.  An interesting development in string theory is the so called AdS/CFT correspondence\cite{maldacena}.  This is a result due to Maldacena, which states that there is a correspondence between string theory in a d+1 dimensional spacetime asymptotic to anti-de-Sitter spacetime (AdS) and a conformal field theory (CFT) on the d dimensional boundary of that spacetime.  In particular, general relativity in a 5 dimensional asymptotically anti-de-Sitter spacetime corresponds to a strongly coupled field theory in ordinary 4 dimensional spacetime.  This is intriguing because it is notoriously difficult to perform calculations in strongly coupled field theories.  But such calculations would be necessary to describe such processes as the outcome of heavy ion collisions or the properties of certain types of superconductors.  One part of the correspondence is that black holes correspond to field theory thermal states.  This means that the approach to thermal equilibrium in a strongly coupled field theory can be modeled using gravitational collapse and black hole formation.  But gravitational collapse is something that has been extensively studied using numerical relativity.  The numerical relativity methods needed modification so that they could be applied in spacetimes that are 5 dimensional and asymptotic to anti-de-Sitter spacetime (rather than the usual 4 dimensional asymptotically flat spacetimes that have usually been treated in numerical relativity).  Such modifications are challenging, but the resulting simulations yield a powerful new method for studying strongly coupled field theories.
	
Section 2 will give a short summary of numerical relativity techniques and uses.  The following sections will cover the applications of numerical relativity to singularities (section 3),  critical gravitational collapse (section 4), and higher dimesions including the AdS/CFT correspondence (section 5).  Conclusions will be given in section 6.

\section{Numerical Relativity}

The equations of physics are differential equations, and numerical techniques are methods that approximate those differential equations in ways that are suitable for simulation on a computer.  The most straightforward of these techniques are those involving finite differences.  We approximate a function $f(x)$ by giving its values on equally spaced points, ${f_i}=f(i \delta)$ where $i$ is an integer and $\delta$ is the spacing.  Then the $f_i$ allow us ways of varying accuracy to approximate derivatives of $f$.  For example ${f'}$ at point $i$ is approximately 
$({f_{i+1}}-{f_i})/\delta$.  However, this approximation makes an error of order $\delta$.  A better approximation for $f'$ is $({f_{i+1}}-{f_{i-1}})/(2\delta)$, for which the error is of order ${\delta ^2}$.  Similarly, the approximation of $f''$ as $({f_{i+1}}+{f_{i-1}}-2{f_i})/({\delta^2})$ has an error of order ${\delta^2}$.

In general relativity, the quantity describing the gravitational field is the metric tensor $g_{\alpha \beta}$, which can be introduced as follows: we collectively denote the space and time coordinates $x^\alpha$ where $\alpha$ takes on the values $0,1,2,3$ with $x^0$ being the time coordinate $t$ and ${x^1}, \, {x^2}, \, {x^3}$ being three spatial coordinates.  We usually think of a trajectory as being ${\vec x}(t)$, that is the spatial coordinates as a function of the time coordinate.  However, we can equivalently think of a trajectory as a parametric curve with all four spacetime coordinates $x^\alpha$ given as a function of some parameter $\lambda$.  Now consider an observer traveling along some trajectory that begins at an initial spacetime point ${x^\alpha _i}$ and ends at a final spacetime point 
${x^\alpha _f}$.  How much time $\Delta \tau$ elapses on a watch carried by that observer?  Before relativity we would have said $\Delta \tau = {t_f}-{t_i}$.  However, general relativity says that
\begin{equation}
\Delta \tau = {\int _{\lambda _i} ^{\lambda _f}} \; d \lambda \; {\sqrt {- {g_{\alpha \beta}} {\frac {d{x^\alpha}} {d \lambda}} {\frac {d{x^\beta}} {d \lambda}} }}
\label{tau}
\end{equation}       
Here we have applied the ``Einstein summation convention'' in which the indices $\alpha$ and $\beta$ are to be summed over.  What makes general relativity a theory of gravity is the statement that objects in free fall behave in such a way as to maximize the elapsed proper time.  That is, the quantity $\Delta \tau$ is the action for falling objects.
An epression identical to eqn. (\ref{tau}) except for the minus sign gives the length of a curve in a curved space, therefore in analogy the quantity $g_{\alpha \beta}$ is called the metric of a curved spacetime.  

Wheeler once summarized general relativity as ``spacetime tells matter how to move, and matter tells spacetime how to curve.''  Using eqn. (\ref{tau}) as an action is the ``spacetime tells matter how to move'' part.  Now we come to the ``matter tells spacetime how to curve'' part, which is unfortunately much more complicated.  This part is expressed in the Einstein field equations, which take the form
\begin{equation}
{G_{\alpha \beta}} = {\frac {8\pi G} {c^4}} {T_{\alpha \beta}} 
\label{efe}
\end{equation}  
Here $G_{\alpha \beta}$ represents the curvature of spacetime, $T_{\alpha \beta}$ represents the energy density of the matter, $c$ is the speed of light, and $G$ is Newton's gravitational constant.  $G_{\alpha \beta}$ is a complicated function of the metric and its first two derivatives.  Nonetheless, since we are thinking in terms of solving the equations by computer simulations, we should not be put off by a little complication.  In particular, the fact that eqn. (\ref{efe}) contains no more than two derivatives immediately suggests the following strategy: perform some algebra on eqn. (\ref{efe}) so that the second time derivative of $g_{\alpha \beta}$ is on the left hand side and all other terms are on the right hand side.  This is then an equation of motion for $g_{\alpha \beta}$ for which the initial data are the initial values of $g_{\alpha \beta}$ and its time derivative.  Replace derivatives with finite differences, write a computer program that implements those finite difference equations, choose initial data, and run the program.

Unfortunately, there is a fundamental difficulty with this strategy, one that puzzled Einstein as he developed general relativity:  in general relativity we are allowed to use any coordinate system we like.  However, physical quantities, like elapsed time on a watch held by a particular observer, cannot depend on our choice of coordinate system.  It then follows from eqn. (\ref{tau}) that when we change coordinates the expressions for $g_{\alpha \beta}$ as a function of those coordinates changes.  Now suppose that the new coordinates differ from the old coordinates only after some time $t_c$ later than the initial time.  Then the two metric expressions will have the same initial data, but different time developments.  Therefore eqn. (\ref{efe}) does not have solutions uniquely determined by initial data, and therefore does not seem to make sense as an equation of motion.

This particular difficulty does not occur in areas of physics other than general relativity; however there is an analogous difficulty for Maxwell's equations.  It is therefore instructive to take a short detour to consider this analogous difficulty and the means of its solution.  Maxwell's equations are written in terms of electric field 
$\vec E$ and magnetic field $\vec B$, but it is helpful to introduce a vector potential $\vec A$ and scalar potential 
$V$ such that ${\vec B} = {\vec \nabla } \times {\vec A}$ and 
${\vec E} = - [{\vec \nabla}V + (1/c) \partial {\vec A}/{\partial t}]$.  This ansatz automatically solves two of the four Maxwell equations, and the remaining two Maxwell equations become equations of motion for $\vec A$ and $V$.  However, it is easy to see that these equations of motion cannot have unique solutions, in particular because the transformation
\begin{equation}
{\vec A} \to {\vec A} + {\vec \nabla}\chi $ , \; \; \;  $V \to V - {\frac 1 c} {\frac {\partial \chi} {\partial t}} 
\label{gauge}
\end{equation}
leaves $\vec E$ and $\vec B$ unchanged.  Eqn. (\ref{gauge}) is called a gauge transformation, and the fact that it doesn't change $\vec E$ and $\vec B$ is called gauge invariance.  The nonuniqueness difficulty is solved by picking a particular gauge.  In particular, it is always possible to choose $\chi$ in eqn. (\ref{gauge}) so that the new 
$\vec A$ and $V$ satisfy the Lorenz gauge condition
\begin{equation}
{\frac 1 c} {\frac {\partial V} {\partial t}} + {\vec \nabla} \cdot {\vec A} = 0
\end{equation}   
With the imposition of the Lorenz gauge condition, Maxwell's equations then become wave equations for ${\vec A}$ and $V$, which then have unique solutions for given initial data.

Is there an analog of Lorenz gauge for general relativity?  It turns out that there is. It is called harmonic coordinates\cite{yvonne} and amounts to choosing coordinates that are solutions of the curved spacetime wave equation.  With this choice of coordinates the Einstein field equations take a form that is a sort of nonlinear wave equation, which has unique solutions given initial data.  One can also choose generalized harmonic coordinates\cite{friedrich} where the coordinates satisfy the curved spacetime wave equation with a specified source.  These generalized harmonic coordinates have been used to simulate the inspiral and merger of black holes.\cite{frans}

More generally, one can write the metric in terms of the spatial metric $g_{ij}$, a scalar $\alpha$ called the lapse, and a spatial  vector $\beta ^i$ called the shift, as follows:
\begin{equation}
{g_{\alpha \beta}}d{x^\alpha}d{x^\beta} =  -{\alpha ^2}d{t^2} + {g_{ij}}(d{x^i}+{\beta^i}dt)(d{x^j}+{\beta^j}dt)
\end{equation}
Here and in what follows we choose units so that the speed of light $c$ is equal to unity.
The lapse and shift can be chosen to be anything one likes (this freedom corresponds to the freedom to choose coordinates) while eqn. (\ref{efe}) provides an equation of motion for the spatial metric.  However, depending on the choice of lapse and shift, the resulting equation of motion may not be hyperbolic, which in turn means that any numerical implementation of the equation is likely to be unstable.  A numerically well behaved system can be obtained by decomposing the spatial metric into an overall conformal factor and a metric of constant determinant.\cite{nakamura,baumgarteshapiro}

To make this general discussion somewhat more concrete, it is helpful to consider the case where the metric is spherically symmetric.  In this case one can choose time $t$ and radial $r$ coordinates so that the metric takes the form
\begin{equation}
{g_{\alpha \beta}}d{x^\alpha}d{x^\beta} = - {\alpha ^2}d{t^2} + {a^2} d {r^2} + {r^2} ( d {\theta ^2} + {\sin ^2} \theta d {\varphi ^2})
\label{sph}
\end{equation}
In geometric terms, the coordinate $r$ is an ``area coordinate'' chosen so that the spheres of symmetry have area
$4\pi {r^2}$ and the time coordinate $t$ is chosen to be orthogonal to $r$.  In general relativity a spherically symmetric gravitational field has no degrees of freedom of its own and is determined by the behavior of the matter. (The corresponding result holds in electromagnetism, where a spherically symmetric electric field is determined entirely by the charge density).  Thus to get any interesting dynamics, we must also treat some form of matter fields.  A simple choice (which is relevant to the treatment of critical collapse in section 3) is a massless scalar field $\phi$, whose equation of motion is the curved space wave equation
${\nabla _\alpha}{\nabla ^\alpha}\phi =0$. With the metric of eqn. (\ref{sph}) the curved space wave equation takes the form
\begin{eqnarray}
{\frac {\partial \Phi} {\partial t}} &=& {\frac \partial {\partial r}} \left ( {\frac \alpha a} \Pi \right )
\label{wave1}
\\
{\frac {\partial \Pi} {\partial t}} &=& {\frac 1 {r^2}} {\frac \partial {\partial r}} \left ( {r^2} {\frac \alpha a} \Phi \right )
\label{wave2}
\end{eqnarray}
Here the fields $\Phi$ and $\Pi$ are given in terms of the scalar field $\phi$ by 
$\Phi = \partial \phi/\partial r$ and $\Pi = (a/\alpha)\partial \phi /\partial t$.
From the Einstein field equations (eqn. (\ref{efe})) it follows that the metric components $\alpha$ and $a$ satisfy
\begin{eqnarray}
{\frac {\partial a} {\partial r}} &=& a \left [ {\frac {1 - {a^2}} {2r}} + r ({\Pi ^2} + {\Phi^2}) \right ]
\label{dra}
\\
{\frac {\partial \alpha} {\partial r}}  &=& \alpha r ({\Pi ^2} + {\Phi^2}) 
\label{dralpha}
\end{eqnarray}
Here units have been chosen so that $4\pi G$ is equal to unity.  Note that these equations are a nice illustration of ``spacetime tells matter how to move and matter tells spacetime how to curve.''  Eqns. (\ref{wave1}-\ref{wave2}) differ from the corresponding equations in ordinary flat spacetime only by the presence of $\alpha$ and $a$ which tell how much the curved spacetime metric of eqn. (\ref{sph}) differs from the flat spacetime metric (in which $\alpha =a=1$).  But $a$ and $\alpha$ are determined by the matter through integrating eqns. (\ref{dra}-\ref{dralpha}).  The constant of integration in these equations are fixed by the conditions that $a=1$ at $r=0$ (which is needed to avoid a singularity at $r=0$) and $\alpha \to 1$ as $r \to \infty$ (which corresponds to the choice that $t$ is the time as measured by a distant observer).    

\section{Singularities}

Stars are in equilibrium, with their self-gravity balanced by the pressure gradients of their matter.  Both the pressure and the self-gravity increase as the star is compressed.  If the star ever gets in a configuration where further compression increases the self-gravity more than it increases the pressure, then equilibrium will no longer be possible: the star will continue to collapse getting ever smaller and denser.  That such a situation is possible was first realized by Chandrasekhar\cite{chandra} and Landau\cite{landau} for the case of white dwarf stars.  General relativity adds further inevitability to this process of complete gravitational collapse: a sufficiently strong gravitational field results in a region where even light cannot escape: a black hole.  Inside a black hole, even standing still would require travelling faster than light, and thus nothing can halt the collapse process, and the matter must collapse completely forming a spacetime singularity.  A mathematical proof that such collapse is inevitable was first provided by Penrose.\cite{penrose}  The conditions of the Penrose theorem are very general, essentially requiring only the existence of a ``trapped surface'' (where gravity is so strong that even the light rays that attempt to be outgoing are pulled inward)  as well as a general condition on matter that essentially states that gravity is always attractive.  However, the conclusion of the theorem is maddeningly non-specific since the ``singularity'' that it requires to exist is simply the statement that some light ray ends.  In particular the theorem does not say that either matter density or spacetime curvature becomes infinite, nor that a black hole forms, nor does it say what the ultimate fate of observers who start out inside the trapped surface is. 

To get a better idea of the nature of spacetime singularities, one should go back to the Einstein field equations (eqn. (\ref{efe})).  If spacetime curvature is blowing up, then one would expect that some terms in the field equations blow up faster than others, and therefore that the approach to the singularity could be well modeled by a simplified set of field equations in which one keeps only the terms that blow up fastest.  This is the approach taken by Belinskii, Khalatnikov, and Lifschitz\cite{bkl} (collectively known as BKL).  Essentially the BKL conjecture is that as the singularity is approached, the terms in the field equations that contain derivatives with respect to time are more important than those that contain derivatives with respect to space.  The BKL conjecture leads to some remarkable conclusions, which can be summarized in the following slogan: ``singularities are spacelike, local, and oscillatory, and matter doesn't matter.''  To see what this slogan means, we will consider the implications of the BKL assumptions: (i) ``spacelike'': a statement like ``time derivatives are more important than space derivatives'' makes sense only after one has chosen the coordinates.  BKL used the amount of time to the singularity as their time coordinate, and thus the singularity is at $t=0$.  A surface of constant time is called ``spacelike'' and one property of a spacelike singularity is that no observer can see the singularity before they hit it.  It was known that the singularity of a Schwarzschild black hole or of Big Bang cosmology is spacelike, but the BKL assumption says that this property holds more generally.  (ii) ``local'': if one can neglect space derivatives then the dynamics is the same as those where the space derivatives are zero, that is a homogeneous (same at all positions in space) spacetime.  But it can be the dynamics of a different homogeneous spacetime at each spatial point.  Thus the dynamics at each spatial point is independent of the dynamics of even nearby spatial points.  (iii) ``oscillatory'': 
since the BKL assumption reduces the dynamics at each spatial point to that of a homogeneous spacetime, we need to know what those dynamics are.  To start with we consider a particular class of homogeneous spacetimes called Kasner spacetimes which have a metric given by
\begin{equation}
{g_{\alpha \beta}}d{x^\alpha}d{x^\beta} = - d{t^2} + {t^{2{p_1}}}d{x^2} + {t^{2{p_2}}}d{y^2} + {t^{2{p_3}}}d{z^2}
\label{Kasner}
\end{equation}
where the $p_i$ are constants.  Here the singularity is at $t=0$.  Note that this expression looks very similar to the Friedmann-Lemaitre-Robertson-Walker (FLRW) metric of cosmology.  In fact, for ${p_1}={p_2}={p_3}=1/2$, eqn. (\ref{Kasner}) gives the usual radiation dominated FLRW metric of the early universe, while for ${p_1}={p_2}={p_3}=2/3$ the metric is matter dominated FLRW.  The metric in eqn. (\ref{Kasner}) satisfies the vacuum Einstein field equations provided that the $p_i$ satisfy the conditions
\begin{equation}
{\sum _i} {p_i} = 1, \; \; \; {\sum _i} {p_i ^2} = 1 .
\label{Kasnervac}
\end{equation} 
In contrast to cosmology where all the $p_i$ are equal, the vacuum conditions of eqn. (\ref{Kasnervac}) require that two of the $p_i$ be positive and one negative.  Thus in the situation of gravitational collapse (i.e. going towards the singularity) two directions are contracting and one is expanding.  Kasner spacetimes are very special homogeneous spacetimes.  However, in a general vacuum homogeneous spacetime the dynamics consists of ``epochs'' where the spacetime is approximately Kasner, punctuated by short ``bounces'' where the metric rapidly changes from one Kasner spacetime to a different one with different values of the $p_i$.  
Thus an object that falls towards a singularity is alternately stretched and squeezed in each direction (though overall squeezed more than stretched).  (iv) ``matter doesn't matter'': `anisotropy' is the name given to the property of different directions having different rates of contraction (or expansion).  Due to the nonlinearity of the Einstein field equations, anisotropy plays a similar role in the equations to energy density.  Thus, one can ask which term is larger as the singularity is approached.  It turns out that the energy density of both ordinary matter and radiation becomes negligible compared to the effective energy density of anisotropy as the singularity is approached.  However, for scalar field matter the energy density grows at the same rate as the anisotropy energy density as the singularity is approached.  Thus if one does not have scalar field matter, then sufficiently close to the singularity one can neglect all matter terms in the field equations and simply treat the vacuum field equations.

BKL checked that their assumptions were consistent with the Einstein field equations; but that doesn't mean that those assumptions hold in general physical situations.  In order to check the BKL conjecture, Berger and Moncrief\cite{bkb1} performed numerical simulations of the approach to the singularity.  For a thorough review of simulations of singularities, the reader is referred to the review paper of \cite{bergerlr} and references therein.  (For a short overview of the subject, as well as the topics of sections 4 and 5, see \cite{thorne}).  Instead we will just touch on some of the main features of this subject.  As with many numerical research programs, it is best to start with simple cases with symmetry and then eventually work up to the general case with no symmetry.  Berger and Moncrief studied a class of spacetimes called the Gowdy spacetimes that have two spatial symmetries.  The Gowdy spacetimes can thus be thought of as collapsing closed universes with circulating plane gravitational waves.  The simulations of \cite{bkb1} showed that the BKL conjecture is correct for these spacetimes: as the singularity is approached, the spatial derivatives play a negligible role in the field equations and one gets dynamics at each spatial point that looks like Kasner punctuated by bounces.  However, Berger and Moncrief also found a phenomenon that had not been anticipated by BKL and that has since come to be known as ``spikes.''  The general dynamics of homogeneous spacetimes consists of a series of bounces between Kasner epochs; however there are exceptional cases where the spacetime stays in a particular Kasner epoch rather than bouncing to the next one.  A spike occurs when a particular spatial point is stuck in the old epoch while its neighbors eventually bounce to the new one.  This gives rise to an ever narrower region of old epoch Kasner surrounded by new epoch Kasner.  Because spikes become arbitrarily narrow as the singularity is approached, they are a challenge to the numerical simulations.  They are also a challenge to the mathematical treatment of the spacetimes.  Nonetheless, despite these challenges, Berger and Moncrief were able to succesfully complete their simulations, and eventually Ringstrom\cite{ringstrom1,ringstrom2} produced a mathematical proof that the Gowdy spacetimes behave in the way shown in the simulations of \cite{bkb1}.  

Berger and Moncrief generalized their methods and simulations to the case with only one symmetry.\cite{bkb2} However, simulations of the general case with no symmetries required a somewhat different set of methods based on a different way of treating the Einstein field equations.  Recall that one of the difficulties with treating singularities is that physical quantities go to infinity there.  However, if two quantities blow up at the same rate, then their ratio does not blow up.  This is the insight behind the method of Uggla et al.\cite{uggla} which writes the Einstein field equations in terms of scale invariant quantities that are obtained by dividing physical quantities by the expansion (essentially the generalization of the Hubble constant of cosmology).  The equations of \cite{uggla} were modified to make them suitable for numerical simulation\cite{dg1,dg2} and used to simulate the approach to the singularity for the general case of no symmetry.  The results of \cite{dg1,dg2} support the BKL conjecture: as the singularity is approached spatial derivatives become less important, and the dynamics at each spatial point consists of a series of Kasner epochs punctuated by short bounces.  Unfortunately, the simulations of \cite{dg1,dg2} did not have enough spatial resolutions to treat the spikes.  However, simulations using the methods of \cite{dg1,dg2} with greatly enhanced resolution provided by parallel computing\cite{meandfrans} are currently under way.  

Note that the BKL picture of the nature of singularities depends on matter being ``sufficiently tame.''  In particular the exotic sorts of matter used in the ekpyrotic cosmological scenario\cite{steinhardt} can lead to the singularity becoming homogenized so that it becomes that of an FLRW cosmology, or even to the singularity being replaced by a bounce to an expanding universe.  For a numerical treatment of these issues see\cite{mefranspaul1,mefranspaul2}.  Quantum gravity effects may also lead to the singularity being replaced by a bounce.  In particular, see \cite{abhay,param} for an analytic and numerical treatment of this issue from the point of view of loop quantum cosmology.  Here the theory itself is naturally formulated in terms of finite difference equations which lend themselves to a numerical treatment.

Though the BKL conjecture, and the numerical simulations that support it, indicate that singularities are spacelike, there is a different set of reasoning that indicates that spacetime singularities are null.  Recall that a null surface is the surface traced out by a set of light rays, and that in particular the event horizon of a black hole is a null surface.  When a star collapses to form a black hole, the exterior of the black hole eventually settles down to a stationary state described by the Kerr metric.  Though the interior of the black hole does not settle down to anything, nonetheless the fact that the exterior settles down to the Kerr metric leads by continuity to the expectation that there well be a region of the interior near the horizon that should be close to the interior Kerr metric.  Inside the black hole event horizon the Kerr metric has an inner horizon which is also a null surface.  However, the inner horizon is unstable, so for a spacetime that begins close to the Kerr metric, the inner horizon should be replaced by a singularity.  There are a variety of analytic arguments, mathematical results, and numerical simulations that indicate that this singularity maintains the inner horizon's character as a null surface.  The Schwarzschild spacetime (describing nonspinning black holes) is spherically symmetric; but the Kerr metric (describing spinning black holes) is only axisymmetric and is therefore more complicated than the Schwarzschild metric.  However, a black hole with charge but no spin is described by the Reissner-N\"ordstrom metric, which is spherically symmetric like Schwarzschild, but has an unstable inner horizon like Kerr.  Much of the treatment of null singularities is done in the Reissner-N\"ordstrom metric, with the hope that any results that obtain in this case will also hold in the Kerr case.  Poisson and Israel\cite{eric} treated ingoing and outgoing streams of fluid in a charged black hole interior and showed that they would result in the formation of a null singularity.  Brady and Smith\cite{brady} performed numerical simulations of an uncharged scalar field in the presence of a pre-existing charged black hole.  Hod and Piran\cite{piran} performed simulations of the collapse of a charged scalar field to form a charged black hole.  In all these cases, the singularity is null.  There is also a mathematical proof for the case of an uncharged scalar field in the presence of a pre-existing charged black hole.\cite{dafermos}  The null singularity axisymmetric vacuum case is not so well treated.  However, there is recent numerical work to first order in perturbation theory around a Kerr background.\cite{burko} There is also some preliminary mathematical work to generalize the result of \cite{dafermos} to the case of a rotating black hole.\cite{luk}

So if there are numerical indications that singularities are spacelike, and numerical indications that singularities are null, then which is it?  Actually, the answer is probably both.  A null surface consists of the paths of light rays, and it remains null only as long as those light rays don't cross.  But matter tends to focus light rays and thus can lead to their crossing.  This reasoning about null surfaces applies to the null singularities expected to form in black hole interiors: when the light rays of the null singularity end it is expected that the null singularity turns into a spacelike singularity described by the BKL approximation.  The interior of a black hole would then contain two types of singularity: a spacelike BKL type singularity encountered by those observers present when the black hole forms, and a null singularity encountered by those observers who fall into the black hole long after it forms.  It would be helpful to have numerical  simulations that could verify (or falsify) this picture.  One difficulty with the existing simulations that show a BKL type singularity is that for numerical convenience they are usually done in a spatially closed spacetime (so that one can apply spatially periodic boundary conditions, which make the numerics simpler).  However, this means that none of these simulations directly describe the behavior of gravitational collapse in an asymptotically flat spacetime (i.e. a spacetime describing an isolated object where gravity gets weak when one gets far from that object).  Since BKL dynamics is local, there should be no obstruction to having a spacetime that is asymptotically flat and has a region with a BKL singularity.  However, it would be helpful to have a simulation in which this occurs.  Preliminary work along these lines is given in \cite{ryo}.                

\section{Critical Gravitational Collapse}

A fundamental issue in general relativity, which goes by the name of ``cosmic censorship,'' is the question of whether the singularities that are produced in gravitational collapse are hidden inside black hole event horizons.  One approach to this question begins with the following basic properties of (nonspinning) black holes: (i) the tidal force outside of a black hole of mass $M$ is proportional to $M/{r^3}$ where $r$ is the area coordinate described in eqn. (\ref{sph}), and (ii) the black hole event horizon is located at $r=2M$. (here we are using units where $G$ and $c$ are unity).  It then follows that the maximum tidal force visible to observers outside the black hole is proportional to $M^{-2}$.  Thus, the smaller the black hole, the larger the visible curvature.  Thus in some sense a ``zero mass black hole'' would have infinite curvature visible to outside observers and would therefore be a naked singularity.  How would one make such a zero mass black hole?  Recall that weakly gravitating objects won't form black holes at all, while very strongly gravitating objects might be expected to make large mass black holes.  If one considers a family of objects depending on a parameter $p$ interpolating between weakly gravitating and strongly gravitating, then there will be a critical parameter $p*$ at which black hole formation first happens.  The question is whether this threshold of black hole formation occurs at finite black hole mass or zero mass.  

To answer this question, Choptuik\cite{choptuik} performed numerical simulations of the collapse of a spherically symmetric scalar field (as described by eqns. (\ref{wave1}-\ref{dralpha})).  The result of \cite{choptuik} is that near the threshold the black hole mass
$M$ behaves as follows:
\begin{equation}
M \propto {{(p-p*)}^\gamma}
\label{bhmass}
\end{equation}    
where $\gamma$ is a universal constant that does not depend on which one parameter family of scalar field initial data one chooses.  Furthermore, the critical solution (the result of evolving the $p=p*$ initial data) has the property of discrete self-similarity: after a certain amount of time the solution repeats itself, with the overall scale of space shrunk.  In other words, for the critical solution the scalar field collapses, but most of it rebounds and radiates away leaving a smaller amount of scalar field which collapses on a smaller time scale and most of it rebounds and radiates away leaving a yet smaller amount of scalar field....   This whole infinite sequence of collapses takes a finite total amount of time, leaving a naked singularity.  Thus it is possible to make a naked singularity, but only for the very special initial data with $p=p*$.  For all other values of $p$, either a black hole forms or the field completely disperses.  Thus the critical solution provides a counter example to one version of cosmic censorship (naked singularities never happen) but not to a weaker version of cosmic censorship (naked singularities do not happen in generic situations).  

There is a connection between the discrete self-similarity of the critical solution and the black hole mass scaling relation of eqn. (\ref{bhmass}).  The critical solution has exactly one unstable mode (which is why we can find it by tuning a single parameter in the initial data).  The mode grows as $e^{\kappa \tau}$ where $\kappa$ is a constant and $\tau$ is the logarithm of the time to the naked singularity.  It then follows that any quantity with dimensions of length will behave like ${|p-{p*}|}^\gamma$ where $\gamma =1/\kappa$.  In particular, since in units where $G=c=1$ black hole mass has units of length, the relation in eqn. (\ref{bhmass}) follows.  Now consider the case of subcritical collapse (i.e. below the threshold of black hole formation).  Then the scalar field collapses and disperses without forming a black hole.  Let $R_{\rm max}$ be the maximum spacetime curvature that occurs in this process.  Then since $R_{\rm max}$ has units of length to the power -2, it follows that in subcritical collapse
\begin{equation}
{R_{\rm max}} \propto {{({p*}-p)}^{-2 \gamma}}
\label{curv}
\end{equation}  
where $\gamma$ is the same constant that occurs in the black hole mass scaling relation.  Numerical simulation of subcritical collapse\cite{meandcomer} verify that eqn. (\ref{curv}) holds.  

Since the results of \cite{choptuik} were obtained using a particular type of matter, a massless scalar field, a natural question to ask is whether one obtains the same result with different types of matter.  And indeed simulations of critical collapse of many different types of matter in spherical symmetry have been done (see the review paper of 
\cite{gundlachlr} and references therein).  A scaling law of the form in eqn. (\ref{bhmass}) continues to hold, however the numerical value of $\gamma$ depends on the type of matter.  In addition, some types of matter have critical solutions that are discretely self-similar, while others have critical solutions that are continuously self-similar.  A more challenging question is whether the critical collapse phenomena continue to hold when one no longer makes the assumption of spherical symmetry.  Recall that the self-similarity of the critical solution means that an examination of critical collapse requires accurate numerical treatment on a very wide range of spatial and temporal scales.  In spherical symmetry one has effectively only one spatial dimension and can therefore obtain a large spatial resolution with a modest allocation of computer memory, especially if (as done in \cite{choptuik}) one also uses Adaptive Mesh Refinement to add extra spatial points where they are needed.  Nonetheless, there have been a few successful simulations of critical collapse in the case of axisymmetry, in particular a simulation of the vacuum case\cite{evans} and of the case with scalar field matter.\cite{mattaxi} 
There has also been a preliminary exploration of the case with no symmetry \cite{pablo} though not with enough resolution for a definitive conclusion.  

The study of critical collapse usually uses specialized numerical methods.  However, since the methods of \cite{frans} and \cite{nakamura,baumgarteshapiro} have been so successful in simulating binary black hole systems, it is natural to see whether those methods would also be suitable for studying critical collapse.  In particular the methods of \cite{frans} have been used in \cite{sorkin1} to study vacuum axisymmetric critical collapse and in \cite{sorkin2} to study critical collapse of a spherically symmetric scalar field.  The methods of \cite{nakamura,baumgarteshapiro} have been used in \cite{matt2} to study critical collapse of a spherically symmetric scalar field, and in \cite{carstenandtom} to study critical collapse of an axisymmetric fluid.

\section{Higher Dimensions and AdS/CFT}

Because string theory requires more than 4 spacetime dimensions, the study of string theory has led to the study of general relativity in more than 4 dimensions.  This is especially the case because one can think of the low energy limit of string theory as general relativity in higher dimensions and with a somewhat exotic collection of matter fields.  One question that immediately suggests itself in such a study is what are the higher dimensional analogues of black holes?  Though many such ``black objects'' have been found\cite{garyandtoby} using a variety of analytical and numerical methods, one simple method for finding higher dimensional black objects comes from the following elementary result of differential geometry: given an $n$ dimensional vacuum spacetime and an $m$ dimensional flat space, the product of these two spaces is an $n+m$ dimensional vacuum spacetime.  In particular, the product of an ordinary 4 dimensional Schwarzschild black hole with a line is a five dimensional ``black string.''  Black holes are of interest for astrophysics not only because they can form in gravitational collapse, but also because they remain after forming, in other words black holes are stable.  Thus for these higher dimensional black objects, one would like to know are they stable? (and in particular one would like to know if the 5 dimensional black string
is stable).  This question was answered by Gregory and Laflamme\cite{glf} who found an unstable mode of the black string.  One way to understand why one might expect black strings to be unstable comes from the subject of black hole thermodynamics: as first shown by Bekenstein\cite{bhentropy1} and Hawking\cite{bhentropy2}, black holes can be thought of as thermodynamic systems, with the black hole area playing the role of entropy.  But entropy tends to increase, thus one might expect an object to be unstable if there is a state of higher entropy that it can evolve to (for a more rigorous version of this argument for instability of black objects, see\cite{wald}).  For a sufficiently long black string, there is a configuration of the same energy but larger entropy (i.e. horizon area) consisting of 5 dimensional black holes.  Thus, as argued in\cite{glf} one might expect a 5 dimensional string to ``pinch off'' into 5 dimensional black holes.  However, there is a disturbing implication of this argument: as shown by Horowitz and Maeda\cite{gary}, a black string that pinches off into black holes must give rise to a naked singularity.  So do unstable black strings evolve to pinch off black holes and naked singularities?  In order to find out, Choptuik et al.\cite{ubc} performed numerical simulations of evolving black strings.  Unfortunately, the method of \cite{ubc} did not allow the simulations to run long enough to settle the question.  However, these methods were improved by Lehner and Pretorius\cite{luisandfrans} who were able to settle the question.  In the simulations of \cite{luisandfrans} the black string evolves to a collection of 5 dimensional black holes connected by lengths of black string of much smaller radius than the original black string.  But these lengths of thinner black string are themselves unstable and evolve to a collection of black holes connected by lengths of even thinner black string, and so on.  The result of ``and so on'' is an infinite collection of black string instabilities that occur on ever shorter time scales, so that the whole process takes a finite amount of time, thus resulting in a naked singularity.  Thus the simulations of \cite{luisandfrans} show that cosmic censorship is violated, and in constrast to the situation of \cite{choptuik} this violatiion is generic.  Thus, it seems that the overall conclusion that we can draw is that cosmic censorship holds in 4 spacetime dimensions, but not in 5 spacetime dimensions.   

One subject of great current interest in the string theory community is the so-called ``AdS/CFT correspondence.''  To introduce this subject, we first consider some of the properties of anti-de Sitter spacetime (AdS), which can be thought of as a homogeneous, isotropic cosmology with a negative cosmological constant and no other matter.  AdS has the strange property that light rays can get to spatial infinity in a finite time.  Thus to make sense of the propagation of light rays in AdS, one must add boundary conditions at spatial infinity.  In the context of string theory, one thinks of $n$ dimensional AdS as being endowed with an $n-1$ dimensional ``boundary at infinity'' and considers conformal field theories (CFT) on that boundary manifold.  Then as first shown by Maldacena\cite{maldacena} there is a correspondence between string theory in AdS at weak coupling and a certain CFT on the boundary at strong coupling.  In particular, it is thought that black holes in $n$ dimensional classical GR with a negative cosmological constant correspond to thermal states in a strongly coupled $n-1$ dimensional CFT.  Thus one is led (even if, like the present author, one is somewhat skeptical about the exact extent, meaning, and interpretation of the ``correspondence'') to consider the properties of black holes and gravitational collapse in spacetimes that are asymptotically AdS.

One of the first questions of this sort that one might want to investigate is what are the properties of critical gravitational collapse in asymptotically AdS spacetimes.  And indeed numerical simulations of this sort were performed by \cite{bizon}.  Since critical collapse occurs on very small spatial scales and thus very large energy densities one might at first expect the same sort of results as in \cite{choptuik}, and indeed one obtains a mass scaling relation of the form of eqn. (\ref{bhmass}) with some critical parameter ${p*}_1$ for black holes formed shortly after the collapse process begins.  However, for parameter $p$ slightly less than ${p*}_1$, the scalar field after dispersing to infinity reflects off of AdS infinity, recollapses and this time forms a black hole.  One then obtains another scaling relation of the form of eqn. (\ref{bhmass}), but now with a different critical parameter ${p*}_2$ for black holes formed after one bounce off of spatial infinity.  This succession of scaling relations repeats itself at ever smaller critical parameter with the result that no matter how small the initial amplitude of the scalar field, eventually a black hole forms: anti de Sitter spacetime is unstable!  In hindsight, there are two reasons to expect such an instability, one from the AdS/CFT correspondence and one from the mathematical study of partial differential equations (PDE).  The AdS/CFT reason: recall that there is supposed to be a correspondence between black holes in AdS and thermal states in the CFT.  Thus the statement that in AdS a black hole forms should correspond to the statement that in the CFT things eventually come to thermal equilibrium.  Since the CFT outcome (things eventually come to thermal equilibrium) is something we should expect, the AdS outcome (eventually a black hole forms) should also be expected.
The PDE reason: in nonlinear theories like general relativity one obtains stability results (e.g. the stability of Minkowski spacetime\cite{ck}) only because waves can disperse.  Since in AdS waves are unable to disperse because they keep getting reflected back by bouncing off of spatial infinity, one should not expect AdS to be stable.  Nonetheless, there do appear to some special configurations of scalar field that are ``islands of stability'' in the sense that they do not lead to the formation of black holes.\cite{lehnerliebling,bizon2,lehnergreen}

Since in quantum field theory there is great interest in finding methods to calculate strong field processes, it is natural to use the AdS/CFT correspondence to find an AdS analog of the field theory system one is interested in (using the so-called ``AdS/CFT dictionary''), perform numerical simulations of the AdS system and translate (again using the ``dictionary'') the results of those simulations into results for the field theory.  There are many such studies, and we refer the reader to the review paper of \cite{nrhep} and references therein for details.  However, we will say a few words about two such systems: (i) heavy ion collisions and (ii) superconductors.  First note that the field theories that describe these processes (quantum chromodynamics (QCD) for heavy ion collisions and electromagnetism coupled to an effective field theory of Cooper pairs of electrons for superconductors) are {\emph {not}} conformal field theories, much less the sort of supersymmetric conformal field theories that are the CFTs one obtains from the AdS/CFT correspondence.  Therefore one will not obtain exact results to be precisely compared to experiments.  The best that one can hope for is to concentrate on phenomena that are fairly universal over a wide class of field theories that contain both the physical field theory one is interested in and the CFT that corresponds to AdS.  Then one might obtain for those phenomena rough qualitative agreement with the experiments.  For heavy ion collisions this was done in \cite{yaffe} by simulating the collision of plane gravitational waves in anti de Sitter space.  For superconductors what is simulated\cite{garysc} is the behavior in AdS of scalar fields propagating in the presence of a charged black hole.  Most of these simulations use the simplifying assumption of either spherical symmetry or plane symmetry.  However, there has also been some work on the more general case of axisymmetry.\cite{bantilan}     

\section{Conclusions}

There has been much numerical work done to investigate the properties of general relativity as a fundamental theory of space and time.  However, there is also much that remains to be done.  Many of the simulations of spacelike singularities have been done for closed universes, so these simulations need to be extended to the asymptotically flat case to verify their relevance for the outcome of gravitational collapse.  The numerical investigation of null singularities is in its infancy and needs to be extended to cases with less symmetry (eventually with no symmetry at all).  Similarly, more work needs to be done on axisymmetric critical gravitational collapse, and eventually (with large amounts of computational resources, possibly leveraged with improved numerical methods) a full treatment of critical collapse with no symmetry.  The black string simulations indicate that cosmic censorship is generically violated in 5 spacetime dimensions.  This important issue needs to be investigated more thoroughly.  In particular, it would be nice to know if one can form a naked singularity from an initial configuration that begins with no pre-existing black object.  Numerical simulations can also be used to bring some order to the zoo of higher dimensional black objects by investigating which of those objects are stable.  Finally, it would be helpful to know with more precision the nature of the correspondence between AdS and CFT (If you posess a copy of the fabled ``AdS-CFT dictionary,'' please  send it to me).  In this way, we could determine {\emph {which}} asymptotically AdS spacetimes have most physical relevance.  We could then study those spacetimes in more detail, using all the numerical relativity methods at our disposal (and if necessary inventing new ones). 

For those of you, dear readers, whose field is not numerical relativity, I hope that this treatment of non astrophysical numerical relativity has been a helpful overview of the subject.  If you wish to learn more about the subject, I encourage you to consult the references to this paper, especially the review papers.\cite{bergerlr,gundlachlr,nrhep}  For those of you whose field {\emph {is}} numerical relativity, I hope you find the description of possible future projects both tempting and useful. (i.e. get to work!)   

\section*{Acknowledgements}
It is a pleasure to thank for many (many!) useful conversations over the years on numerical relativity and related subjects Beverly Berger, Matt Choptuik, Luis Lehner, and Frans Pretorius.  This work was supported in part by NSF grant PHY-1505565 to Oakland University.

\end{document}